\documentclass[12pt]{iopart}
\usepackage{epsfig}
\def\simgt{\rlap{\lower 3.5 pt \hbox{$\mathchar \sim$}} \raise 1pt \hbox {$>$}}
\def\simlt{\rlap{\lower 3.5 pt \hbox{$\mathchar \sim$}} \raise 1pt \hbox {$<$}}
\catcode`\@=11
\def\@citex[#1]#2{\if@filesw\immediate\write\@auxout{\string\citation{#2}}\fi
  \def\@citea{}\@cite{\@for\@citeb:=#2\do
    {\@citea\def\@citea{,\penalty\@m}\@ifundefined
       {b@\@citeb}{{\bf ?}\@warning
       {Citation `\@citeb' on page \thepage \space undefined}}%
\hbox{\csname b@\@citeb\endcsname}}}{#1}}
\def\citer{\@ifnextchar [{\@tempswatrue\@citexr}{\@tempswafalse\@citexr[]}}
\begin{document}

\begin{flushright}
Edinburgh 2000/05 \\
NIKHEF/99-033 \\
hep-ph/0002112
\end{flushright}
\jl{4}

\title[]{Heavy Flavour Production in Two-Photon Collisions}

\author{S Frixione\dag\,  M Kr\"amer\S\ and E Laenen\P}
\address{\dag\ CERN, Theoretical Physics Division, CH-1211 Geneva 23, 
                Switzerland}
\address{\S\ Department of Physics and Astronomy, University of 
 Edinburgh, Edinburgh EH9 3JZ, Scotland}
\address{\P\ NIKHEF Theory Group, Kruislaan 409, 1098SJ, Amsterdam, 
The Netherlands}

\begin{abstract}
We review the production of charm and bottom quarks in two-photon
collisions at $e^+e^-$ colliders. The next-to-leading order QCD
predictions for total cross sections and differential distributions
are compared with recent experimental results.
\end{abstract}

\section{Introduction}
The production of charm and bottom quarks in two-photon collisions at
high-energy $e^+e^-$ colliders provides new possibilities to study the
dynamics of heavy quark production and complements the extensive
analyses that have been carried out at fixed-target experiments and at
other colliders~\cite{FMNR-97}. In two-photon collisions, each of the
photons can behave as either a point-like or a hadronic
particle. Consequently, one distinguishes in such collisions direct-
(both photons are point-like), single resolved- (one photon is
point-like, the other hadron-like), and double resolved (both photons
are hadron-like) production channels. The resolved channels require
the use of parton densities in the photon, whereas the production via
the direct channel is free of such phenomenological inputs.

The mass of the heavy quark, $m_{\rm Q}\gg\Lambda_{\rm QCD}$, sets the
hard scale for the perturbative calculation at small transverse
momentum. It is thus possible to define an all-order infrared-safe
cross section for open heavy flavour production. The heavy quark mass
also ensures that the separation into direct and resolved production
channels is unambiguous at next-to-leading order (NLO). Beyond NLO,
however, the different channels mix and the distinction between the
direct and resolved contributions becomes non-physical and
scheme-dependent.

\section{Results}
Total cross sections and various distributions for inclusive charm and
bottom quark production in two-photon collisions at $e^+e^-$ colliders
have been calculated in \cite{DKZZ-93}, including NLO QCD corrections
for the leading subprocesses.\footnote{The effects of summing large
logarithms $\log (p_T/m_{\rm Q})$ at $p_T \gg m_{\rm Q}$ have been
studied in \cite{CGKKKS-96} at NLO.} At low energies, in the
PETRA/PEP/TRISTAN range, the direct production mechanism by far
dominates the total cross section. At LEP2 energies, the
resolved-$\gamma$ contribution becomes sizeable, up to about $50\%$ of
the total cross section, depending in detail on the choice for the
parton densities in the photon. The cross sections for charmed
particle production are large, giving a total of roughly $200000$
events for an integrated luminosity of $\int {\cal L} =
200~\mbox{pb}^{-1}$ at LEP2; $b$ quark production is suppressed by
more than two orders of magnitude, a consequence of the smaller bottom
electric charge and the phase space reduction by the larger $b$
mass. The inclusion of QCD corrections is important, increasing the
cross section by about $30\%$.

\begin{figure}[p]
\centerline{
   \epsfig{figure=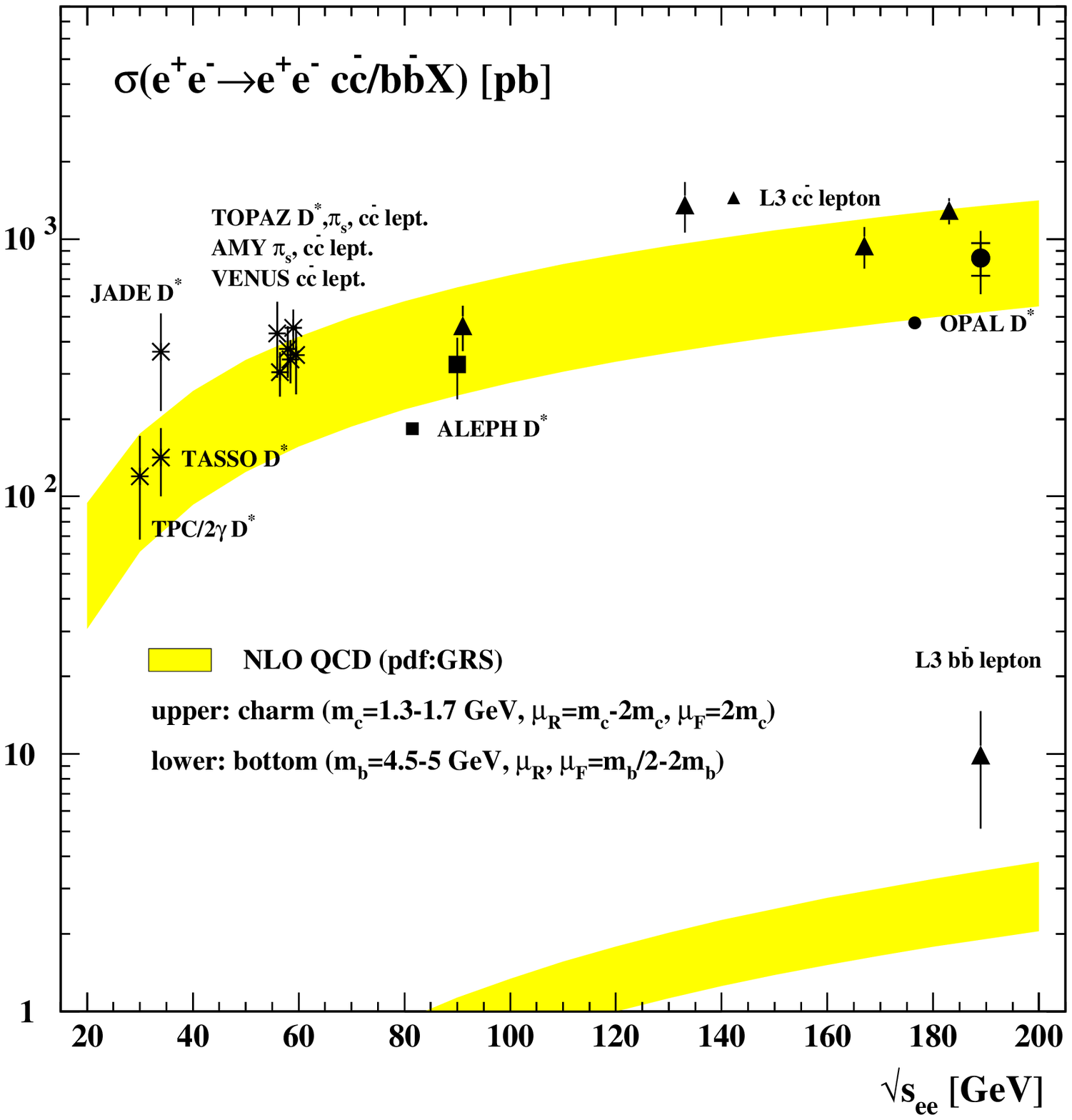,width=1.1\textwidth,clip=}}
\caption[]{ \label{fig:totalxs}
Comparison of NLO QCD predictions \cite{DKZZ-93} and experimental
results \cite{totalxs-exp,diff-exp} for the total charm and bottom
cross section as a function of the $e^+e^-$ collider energy. GRS
photonic parton densities \cite{GRS-99}.}
\end{figure}

In Figure~\ref{fig:totalxs} we compare the NLO predictions for the
total charm and bottom cross section with experimental data
\cite{totalxs-exp,diff-exp} from PETRA energies up to LEP2 energies. The
theoretical predictions appear under control, the uncertainty due to
variation of the heavy quark mass and the renormalization and
factorization scale being approximately $\pm 40\%$, as indicated in
the figure. The overall agreement between the experimental charm data
and the QCD predictions is good, provided higher-order corrections and
resolved-photon contributions are included in the theory. The
experimental and theoretical errors, however, do not allow to
discriminate between different recent sets of photonic parton
distributions \cite{Drees:1990zw}. More data are needed before any
conclusions on two-photon production of bottom quarks can be drawn.

More detailed comparisons between QCD predictions and experimental
data can be performed by using fully differential NLO Monte Carlo
programs \cite{FKL-99}, which have recently been constructed. In these
codes, all final-state kinematical quantities are available on an
event-by-event basis, and it is thus possible to calculate more
exclusive observables at NLO and to include
heavy-quark--to--heavy-meson fragmentation functions.

\begin{figure}[p]
\centerline{
   \epsfig{figure=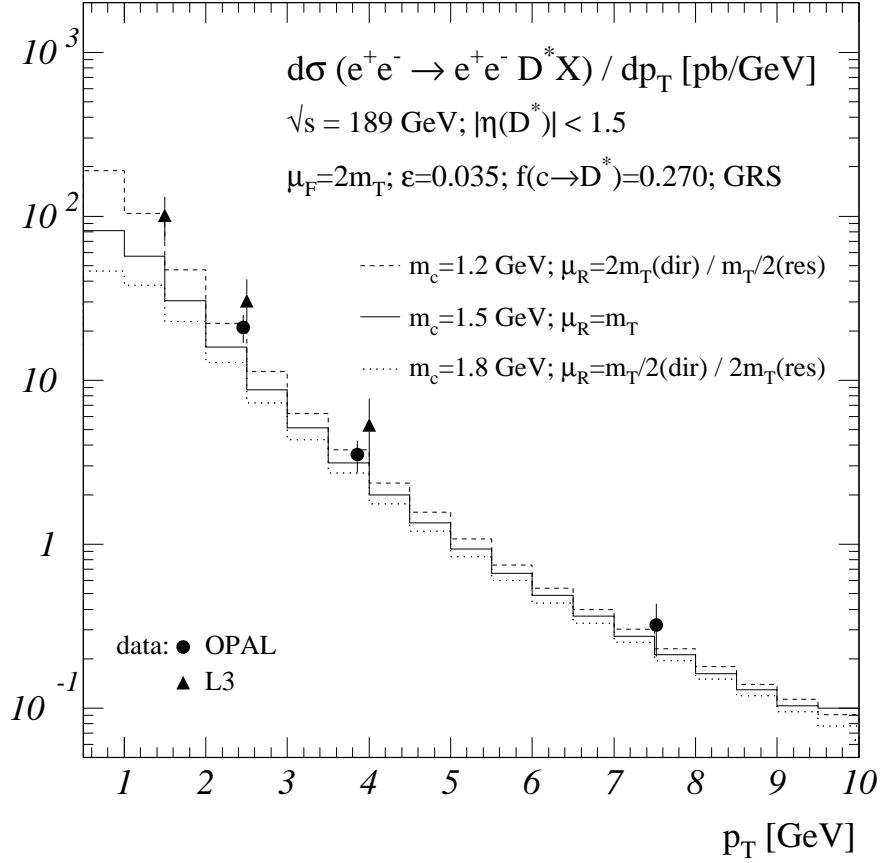,width=0.8\textwidth,clip=}}
\caption[]{ \label{fig:comp-data1}
Comparison between the NLO QCD prediction \cite{FKL-99} and the
OPAL/L3 data \cite{diff-exp} for the $D^*$ transverse momentum
distribution.  The Peterson {\it et al.} fragmentation function
\cite{PSSZ-83} with $\epsilon=0.035$ has been adopted and the probability
for a charm quark to fragment into a $D^*$ meson is set to $f(c\to
D^{*})=0.270$. GRS photonic parton densities \cite{GRS-99}.  The L3
data \cite{diff-exp} have been
reanalyzed using an anti-tag condition for the scattered electrons
\cite{L3-priv}. In order to allow a common
display of the OPAL and L3 data sets, we have scaled the L3 data to
account for the different pseudorapidity range and electron anti-tag
condition; numerically, the effect is $\simlt\;10\%$ and negligible on
the scale of the Figure.}
\end{figure}

The OPAL and L3 collaborations have presented \cite{diff-exp} new data
for $D^*$ production in two-photon collisions, at (mostly)
$\sqrt{s_{e^+e^-}} = 189$ GeV.  Besides the total cross section
$\sigma_{\gamma\gamma}^{D^*}$, both experiments have measured the
differential rate with respect to the $D^*$ transverse momentum,
$d\sigma_{\gamma\gamma}^{D^*} / d p_T^{D^*}$, and pseudorapidity
$d\sigma_{\gamma\gamma}^{D^*} / d \eta^{D^*}$.  In
Figure~\ref{fig:comp-data1} we compare our predictions for the $D^*$
transverse momentum distribution with the OPAL and L3 measurements.
Three different theoretical curves are shown, where the charm quark
mass and the renormalization scale are varied as indicated in the
figure. The shape of the distribution is described well by NLO theory,
while there is a small discrepancy in absolute normalization, in
particular for the L3 data, when central values for the theoretical
input parameters are adopted. A definite statement, however, will only
be possible after the statistical errors affecting the measurements
have decreased. For both experiments, the pseudorapidity distribution
is observed to be essentially flat in the central region, in
accordance with the theoretical expectations \cite{DKZZ-93,FKL-99}.

\section{Summary}
Total and differential heavy-quark production rates in two-photon
collisions at $e^+e^-$ colliders have been studied including NLO QCD
corrections. Compared with charm and bottom production in
hadron--hadron or photon--hadron collisions, the two-photon cross
section appears to be under better theoretical control, mainly because
of the dominance of the direct channel, where QCD uncertainties are
smaller than in the case of the resolved contributions. The overall
agreement between the total cross section measurements and the QCD
predictions is good, provided higher-order corrections and
resolved-photon contributions are included in the theory. We also find
good agreement for the shape of the differential distributions.  The
absolute normalization of the data in the experimentally visible
region is, however, slightly underestimated by NLO theory when central
values for the input parameters are adopted. A special tuning of the
input parameters is needed to improve the agreement, a pattern already
known from other types of collider experiments. When the statistical
significance of the measurements will be increased, heavy quark
production in two-photon collisions will be a valuable tool in testing
the underlying production dynamics.

\ack
We would like to thank Valeri Andreev, Alex Finch, Jochen Patt and
Stefan S\"oldner-Rembold for their encouragement and for valuable
discussions. The work of S.F.\ and M.K.\ is supported in part by the
EU Fourth Framework Programme `Training and Mobility of Researchers',
Network `Quantum Chromodynamics and the Deep Structure of Elementary
Particles', contract FMRX-CT98-0194 (DG 12 - MIHT). The work of E.L.\
is part of the research program of the Foundation for Fundamental
Research of Matter (FOM) and the National Organization for Scientific
Research (NWO).

\section*{References}

\end{document}